\documentstyle[12pt,epsf]{article}

\renewcommand{\thefootnote}{\fnsymbol{footnote}}

\newcommand{\bea}{\begin{eqnarray}}
\newcommand{\eea}{\end{eqnarray}}
\newcommand{\beq}{\begin{equation}}
\newcommand{\eeq}{\end{equation}}
\newcommand{\ewxy}[2]{\setlength{\epsfxsize}{#2}\epsfbox[45 240 320 350]{#1}}

\begin{document}

%%%%%%%%%%% Titlepage

\begin{titlepage}
\begin{flushright}
LPTHE-Orsay 98/37\\
Rome1 1212/98
% \\ hep-ph/9807046
\end{flushright}
\vskip 0.5cm
\begin{center}
{\Large \bf Non-perturbatively Renormalized Light-Quark Masses with the 
Alpha Action}
\vskip1cm 
{\large\bf D.~Becirevic$^a$, Ph.~Boucaud$^a$, J.P.~Leroy$^a$}\\
{\large\bf V.~Lubicz$^b$, G.~Martinelli$^c$, F.~Mescia$^c$}\\

\vspace{.5cm}
{\normalsize {\sl $ ^a$Laboratoire de Physique Th\'eorique et Hautes Energies\\
Universit\'e de Paris XI, B\^atiment 211, 91405 Orsay Cedex, France \\
\vspace{.2cm}
\vspace{.2cm}
$^b$ Dip. di Fisica, Univ. di Roma Tre and INFN,\\
Sezione di Roma Tre, Via della Vasca Navale 84, I-00146 Rome, Italy\\
\vspace{.2cm}
$^c$ Dip. di Fisica, Univ. di Roma ``La Sapienza" and INFN,\\
Sezione di Roma, P.le A. Moro, I-00185 Rome, Italy.}\\
\vskip1.0cm
{\large\bf Abstract:\\[10pt]} \parbox[t]{\textwidth}{ 
We have computed the light quark masses using the ${\cal O}(a^2)$ improved
Alpha action, in the quenched approximation. The renormalized masses have
been obtained non-perturbatively. By eliminating the systematic error coming 
from the truncation of the perturbative series, our procedure removes the 
discrepancies, observed in previous calculations, between the results obtained 
using the vector and the axial-vector Ward identities. It also gives values of
the quark masses larger than those obtained by computing the renormalization 
constants using (boosted) perturbation theory. Our main results, in the
RI (MOM) scheme and  at a renormalization scale $\mu=2$ GeV,  are $m^{RI}_s= 
138(15)$ MeV and $m^{RI}_l= 5.6(5)$ MeV, where $m^{RI}_s$ is the mass of the 
strange quark and $m^{RI}_l=(m^{RI}_u+m^{RI}_d)/2$  the average mass of the 
up-down quarks. From these results, which have been obtained 
non-perturbatively, by using continuum perturbation theory we derive the 
$\overline{MS}$ masses,  at the same scale, and the renormalization group 
invariant ($m^{\scriptsize{RGI}}$) masses. We find $m^{\scriptsize{NLO \ 
\overline{MS}}}_s= 121(13)$ MeV and $m^{\scriptsize{NLO \ \overline{MS}}}_l= 
4.9(4)$ MeV  at the next-to-leading order; $m^{\scriptsize{N}^2\scriptsize{LO 
\ \overline{MS}}}_s= 111(12)$ MeV, $m^{\scriptsize{N}^2\scriptsize{LO \ 
\overline{MS}}}_l= 4.5(4)$ MeV,    $m_s^{\scriptsize{RGI}}=  177(19)$ MeV    
and $m^{\scriptsize{RGI}}_l=  7.2(6)$ MeV at the next-to-next-to-leading 
order. 
}}
\end{center}
\vskip0.5cm 
\centerline{{\em Submitted to Physics Letters B}}
\unboldmath
\end{titlepage}

\renewcommand{\thefootnote}{\arabic{footnote}}
\setcounter{footnote}{0}
\setcounter{equation}{0}
\section{Introduction}
The values of the quark masses  obtained from  lattice simulations
have recently  attracted  the attention in the physics community.
The reason is that, for these quantities,
the lattice method is unique: from the non-perturbative
calculation of hadronic quantities, this approach allows in fact
a consistent determination of the  quark masses, defined as effective
couplings renormalized at short distances. Yet, 
with the errors quoted by the authors ~\cite{allton}--\cite{petronzio},
the values of the
light and strange quark masses computed in different numerical simulations 
(mainly in the quenched approximation) are often in disagreement.  
The differences originate mainly
from two sources: on the one hand, from the different procedures used to 
compute the
renormalized mass from the bare lattice one; on the other, from the different
methods used to extrapolate results, obtained at finite values of $a$, 
to the continuum limit. 
\par Our numerical calculations have been done with the non-perturbatively
improved fer\-mion action (which we will denote
 as the Alpha action) introduced in ref.~\cite{a00},  see also 
 \cite{a0}--\cite{a8}. 
The use of  a non-perturbatively improved action  and operators reduces
discretization errors to ${\cal O}(a^2)$. 
Thus, at least for light quarks
(namely the $u$, $d$ and $s$ quarks), we expect  these errors to be rather 
small, i.e.
much smaller than other systematic effects. The other source of uncertainty
arises from the truncation of the perturbative series 
in the definition of the renormalized mass. This problem
can be eliminated by using a non-perturbative  method
for renormalizing the lattice operators. In our study, we have computed
the relevant renormalization constants $Z_S$  
and $Z_P$, in the chiral limit, using the non-perturbative renormalization 
procedure on quark states proposed in ref.~\cite{g3} (in the following we 
denote this method as NPM).
As a check of the accuracy of the NPM, we have  also computed the 
renormalization 
constants of the vector and axial-vector currents and compared the results 
with 
those obtained in refs.~\cite{a1}, see also \cite{a0}.  At $\beta=6.0$ and 
$6.2$, our results 
for these currents agree within less than $5 \%$
with previous determinations. This makes us confident that the values of $Z_S$ 
and $Z_P$ determined  with the NPM are correct. Note  that, at the same 
values of $\beta$,  the  
value of   $Z_P$ ($Z_S$)  computed in boosted perturbation theory 
is larger than  our estimate by about 
$30 \%$ (by about $10 \%$)~\footnote{ We used for the boosted 
coupling~\cite{parisi,lpm}
$g^2=g_0^2/\langle P \rangle$, where the average plaquette at $\beta=6.2$
($\beta=6.0$) is given
$\langle P \rangle=0.6136$ ($\langle P \rangle=0.5937$) as inferred by 
our simulations.}. This implies that the
use of perturbation theory leads to an underestimate of the quark mass by
an amount  much larger than the expected  discretization 
errors~\footnote{ The observation that the NPM gives larger quark masses 
was first made in
ref.~\cite{allton}, and then confirmed in a systematic study at different 
values
of $\beta$ and with different actions in ref.~\cite{gimenez}.}. 
It also makes  the two determinations of the masses, from the vector and axial 
vector Ward identities, different.
With the NPM, instead, we find essentially the same value of the quark masses
with the two methods. 
\par This letter is focused on the calculation of the
light quark masses: we describe the procedure followed 
in the determination of  the bare  masses, $\tilde m$, 
discuss the definition of  the renormalized and 
of the renormalization group invariant (RGI) ones, 
$\hat m$ and $m^{\scriptsize{RGI}}$, and 
present the final results and errors for these quantities.  
All  details of the (standard) analysis of the light 
hadron spectrum  and decay constants, together with a study of the  
energy-momentum relation,  can be found 
in  ref.~\cite{next}. For the light hadron spectrum, the calibration
of the lattice spacing, the values of the  Wilson parameter
in the chiral limit, $\kappa _{crit}$, and the values of $\kappa$ 
corresponding
to the light (up and down) and strange quark masses, we substantially
agree with previous studies of the same quantities~\cite{gockeler,petronzio}.
We performed  two independent simulations:
a test run  at $\beta=6.0$ on a $16^3 \times 32$ lattice (with a sample
of $45$ configurations at four values of the light quark masses) and  
a run at $\beta=6.2$ on a  $24^3 \times 64$ lattice (on a sample of $100$ 
configurations
at four values of the light quark masses).
The final results are based on the ``data" at $\beta=6.2$. Those
of the test run have only been used as a consistency check of the 
stability of the physical values of the quark masses.
We will compare the results at $\beta=6.0$ with those at $\beta=6.2$ 
at the end of the  paper.
\section{Definition of the quark masses}
\par We now explain the procedure followed to extract the 
physical values of the quark masses.
The starting point 
are the vector and axial-vector Ward identities (WI)~\cite{boc,maiani}
\beq \nabla_\mu \langle \alpha  \hat V_\mu \vert \beta \rangle
= (\hat m_1(\mu) - \hat m_2(\mu) ) \langle \alpha \vert \hat S(\mu) \vert 
\beta \rangle 
\ ,\label{eq:wiv} \eeq
\beq 
\nabla_\mu \langle \alpha \vert \hat A_\mu \vert \beta \rangle
= (\hat m_1(\mu) + \hat m_2(\mu) ) \langle \alpha \vert \hat P(\mu) 
\vert \beta \rangle \ , \label{eq:wia} \eeq
where $\vert \alpha \rangle$ and $\vert \beta \rangle$ represent generic 
external
physical states; $\hat V_\mu$ and $\hat A_\mu$ are the normalized 
currents which  obey  the current algebra commutation relations; $\hat 
S(\mu)$, $\hat P(\mu)$ and
$\hat m_i(\mu)$ are   operators and quark masses renormalized  
at the scale $\mu$ in a given  scheme.  Note that the products
$\hat m(\mu) \hat S(\mu)$
and $\hat m(\mu) \hat P(\mu)$ are regularization, renormalization 
and scale independent. 
\par Ward identities as (\ref{eq:wiv}) and  (\ref{eq:wia})
can also be written  for the lattice currents.
In this case  they are valid  up to terms of ${\cal O}(a)$,
${\cal O}(\alpha_s a)$
or ${\cal O}(a^2)$, depending whether we use the Wilson, the
tree-level improved~\cite{sw} or
non-perturbatively improved  actions and operators~\footnote{
On the lattice $\nabla_\mu$ is written in terms of (improved) 
finite differences.}. 
\par
In order to determine the renormalized masses it is obvious that we have
first  to  specify  the  scheme  used to renormalize the scalar
and pseudoscalar densities.  Given the uncertainties of lattice perturbation
theory,  we have renormalized these quantities by
imposing,  non-perturbatively,  the following  renormalization
conditions~\cite{g3}
\bea \langle q(p) \vert \hat S(\mu) \vert q(p) \rangle 
\equiv   Z_S(\mu) \langle q(p) \vert  S \vert q(p) \rangle &=& 1 , \nonumber \\
\langle q(p) \vert \hat P(\mu) \vert q(p) \rangle \equiv
Z_P(\mu) \langle q(p) \vert  P \vert q(p) \rangle &=& 1 , \label{eq:rc} \eea
in the chiral limit. In the above equations
$\vert q(p) \rangle $ represents an external off-shell quark state with
virtuality $p^2=\mu^2$, and 
the matrix elements are evaluated in the Landau gauge.
The conditions (\ref{eq:rc}) at large values of $\mu^2$  ensure that 
$\hat S$ and $\hat P$ belong to the same chiral
multiplet, i.e. they obey to the relevant Ward identities.
It can be shown  that this is true also at the improved level, i.e.
that,  using the improved action of ref.~\cite{a1}, the  matrix elements 
of $\hat S$ and $\hat P$ renormalized as 
in eq.~(\ref{eq:rc}) have discretization errors of ${\cal
O}(a^2)$~\cite{sharpenp}.
\par Since vector symmetries are preserved by the lattice 
 regularization, unlike axial vector ones, the procedures which are usually
employed to extract the quark masses using the vector or axial vector
Ward identities are different and we discuss them separately.
\begin{itemize}
\item {\bf Determination of  quark masses from the
vector Ward identity.} \par 
>From the lattice version of eq.~(\ref{eq:wiv}), one finds
\beq \hat m(\mu) =  Z^{-1}_S(\mu, m a, 0) \frac{1}{2 a} \left(
\frac{1}{\kappa} -\frac{1}{\kappa_{crit}} \right) \ , \label{eq:massv} \eeq
where,  in the definition of $Z_S$,  we  added those terms which are
necessary to improve the scalar density out of the chiral limit
\beq Z_S(\mu, m_1 a , m_2 a) = Z_S(\mu) \Bigl( 1 + \frac{b_S}{2} (m_1 a + 
m_2 a) \Bigr)\ .
\eeq
By defining
 $Z_S(\mu)= Z^{-1}_m(\mu)$ and $b_S=-2 b_m$, from eq.~(\ref{eq:massv})
we derive  the standard relation
\beq \hat m(\mu) \equiv Z_m(\mu)\,  \tilde m 
= Z_m(\mu) \left[ \frac{1}{2 a} \left(
\frac{1}{\kappa} -\frac{1}{\kappa_{crit}} \right) \left( 1 + b_m \frac{1}
{2 a} \left(\frac{1}{\kappa} -\frac{1}{\kappa_{crit}} \right) \right) 
\right] \ . \label{eq:massav} 
\eeq 
We have used the above equation to determine the quark mass, with 
$Z_m(\mu)=Z_S(\mu)^{-1}$ as computed from eq.~(\ref{eq:rc}), $\kappa _{crit}$ 
as 
fixed  from the squared pseudoscalar meson mass, the calibration of
the lattice spacing either from $m_\rho$ or from $m_{K^*}$ and $1/\kappa=1/
\kappa_{u,d,s}$ 
from the meson spectroscopy. The discussion of the values of $Z_S$ and $Z_P$ 
and the relative errors can be found in sec.~\ref{sec:rqm}.
\par As far as $b_m$  is concerned, the method of ref.~\cite{sharpenp} has 
difficulties in computing 
it.  The gauge invariant procedure of ref.~\cite{g1} has not been
applied to date. A different  approach  to fix $b_m$  non-perturbatively   
has been proposed in 
ref.~\cite{divitiis}. We have tried the same technique and found that the 
results
for $b_m$ (and for all the analogous quantities such as $b_A$, $b_V$ etc.)
are very unstable and that these constants  cannot be determined 
reliably~\cite{mescia2}~\footnote{A new method for a non-perturbative 
determination of the $b$s, based on the lattice chiral Ward identities, 
has been recently presented at the Lattice '98 conference~\cite{bgl}. 
However, this method has not yet been implemented with the non-perturbatively 
improved Alpha action.}.
For this reason, we  preferred to take $b_m$ from boosted perturbation theory, 
using $b_m = b_m^{(0)} + b_m^{(1)} g^2 = -1/2 - 0.0962 \, g^2 = -0.652$ 
\cite{a7}. If the uncertainty on the perturbative value of $b_m$ were 
as large as  $b_m^{(1)} g^2$, this would induce a relative error of 
 ${\cal O}(6\cdot 10^{-4})$ and ${\cal O}(4\cdot 10^{-3})$ on 
$(\tilde m_d + \tilde m_u)/2$ and $\tilde m_s$, respectively. Given the other 
sources of errors
(due to the uncertainty in the determination of  $Z_S$ and to the 
quenched approximation), this
error is negligible and will be ignored in the following.
\par 
In ref.~\cite{gupta}, they  call HS (from Hadron Spectrum)
the method based on eq.~(\ref{eq:massav}). This is rather misleading
because 
the possibility of relating $1/2 a (1/\kappa -1/\kappa_{crit})$, or any 
improved
version of it, to the quark mass, entirely relies  on the validity of the 
lattice 
vector WI.   The hadron spectrum enters in this case,  as in the case
of the axial-vector WI, only because it is needed to fix the bare quark mass
from the physical value of some hadronic quantity. 
\item {\bf Determination of quark masses from the
axial-vector Ward identity.}
\par In this case, the simplest procedure is to use the axial Ward
identity  computed on hadron states at rest,
for  degenerate quark masses. One gets
\bea \hat m(\mu) &\equiv& \frac{Z_A}{Z_P(\mu)} \, \tilde \rho
=  \frac{Z_A(ma,ma)}{Z_P(\mu,ma,ma)}
\frac{\langle \alpha \vert \nabla_0 A_0 + a c_A \nabla^2_0 P \vert \beta 
\rangle}{ 
2 a \langle \alpha \vert P \vert \beta \rangle} \nonumber \\
&=& \frac{Z_A}{Z_P(\mu)}\left[ 
\frac{1 + b_A ma}{1 +  b_P ma} 
\frac{\langle \alpha \vert \nabla_0 A_0 + a c_A 
\nabla^2_0 P \vert \beta \rangle}{ 
2 a \langle \alpha \vert P \vert \beta \rangle} \right] \label{eq:massaa} 
\ ,\eea
where $b_A$ and $b_P$  play the same role for the axial current and the 
pseudoscalar density as $b_S$ in the case of the scalar operator. In the 
absence of a precise non-perturbative determination of these quantities, we 
have used, as we did above for $b_m$,  boosted
perturbation theory, namely $b_A= 1 + 0.1522  g^2=1.240$ and $b_P=1 + 0.1531 
g^2=1.241$ \cite{a7}. The same considerations made for $b_m$ about the 
systematic error induced by perturbation theory in the calculation
of $b_A$ and $b_P$ on the masses
of the light quarks, remain true in this case. 
We extracted the quark mass by computing the ratio in  eq.~(\ref{eq:massaa}) at
values of $\kappa$ corresponding to the physical meson masses ($m_\pi$, $m_K$
and $m_\phi$),
taking the non-perturbative determinations of $Z_A$ and $Z_P(\mu)$, 
evaluated in the  chiral limit. \end{itemize}
\section{Non-perturbatively renormalized quark masses}
\label{sec:rqm}
In this section, we first discuss the values and errors of the renormalization 
constants $Z_S(\mu)$, $Z_P(\mu)$ and $Z_A$.  We then explain  the procedure
followed to extract the bare masses,   $\tilde m$ and $\tilde
\rho$, using eqs.~(\ref{eq:massav}) and  (\ref{eq:massaa}).
>From the $Z$s  and from the values of the bare masses, we have obtained the  
values of the non-perturbatively renormalized quark
masses  in the $RI$ scheme.  Using perturbation theory,  we then get
 the $\overline{MS}$ and RGI masses at the $NLO$ and $N^2LO$.
\par The  relevant Green functions (vertices and propagators) necessary
to the determination of the $Z$s with the NPM have been computed 
with external quark states  of virtuality $\mu$
on the same configurations, and for the same values of $\kappa$, 
as all the other quantities discussed in this paper.  We used 
quark propagators  improved following the strategy of 
refs.~\cite{sharpenp,mescia2}. The Green functions have then been 
extrapolated (linearly in the quark masses) to the chiral limit. 
\par As explained in ref.~\cite{g3},
the NPM is expected to work when $\mu$ satisfies the condition
$\Lambda_{QCD} \ll \mu \ll 1/a$. In this region,  $Z_A$ and $Z_P/Z_S$ should be
independent of $\mu$, since
the non-perturbative method is equivalent to the Ward identities.
We monitored the behaviour of $Z_A$ and $Z_P/Z_S$ as
a function of $\mu$ to find the range of $\mu$ where these quantities exhibit  
a plateau.  At $\beta=6.2$,
we find that $Z_A$ is essentially a constant for $\mu ^2 a^2 \le 2$,
while $Z_P/Z_S$ has a plateau for $1 \le \mu^2 a^2 \le 2$, corresponding to 
$2.8$ GeV $\le \mu  \le 4.0$ GeV. 
>From the analysis of the plateaus of $Z_A$ and $Z_P/Z_S$, we have obtained
$Z_A=0.793(5)$, in agreement with the value $Z_A=0.809$ 
of ref.~\cite{a1},  and $Z_P/Z_S=0.78(1)$. The latter in disagreement with 
the boosted result $Z_P/Z_S=0.95$.  Similar results, although with larger 
statistical
fluctuations, and systematic uncertainties,
have been found at  $\beta=6.0$~\cite{mescia2}.
\par
\begin{figure}[t]
\ewxy{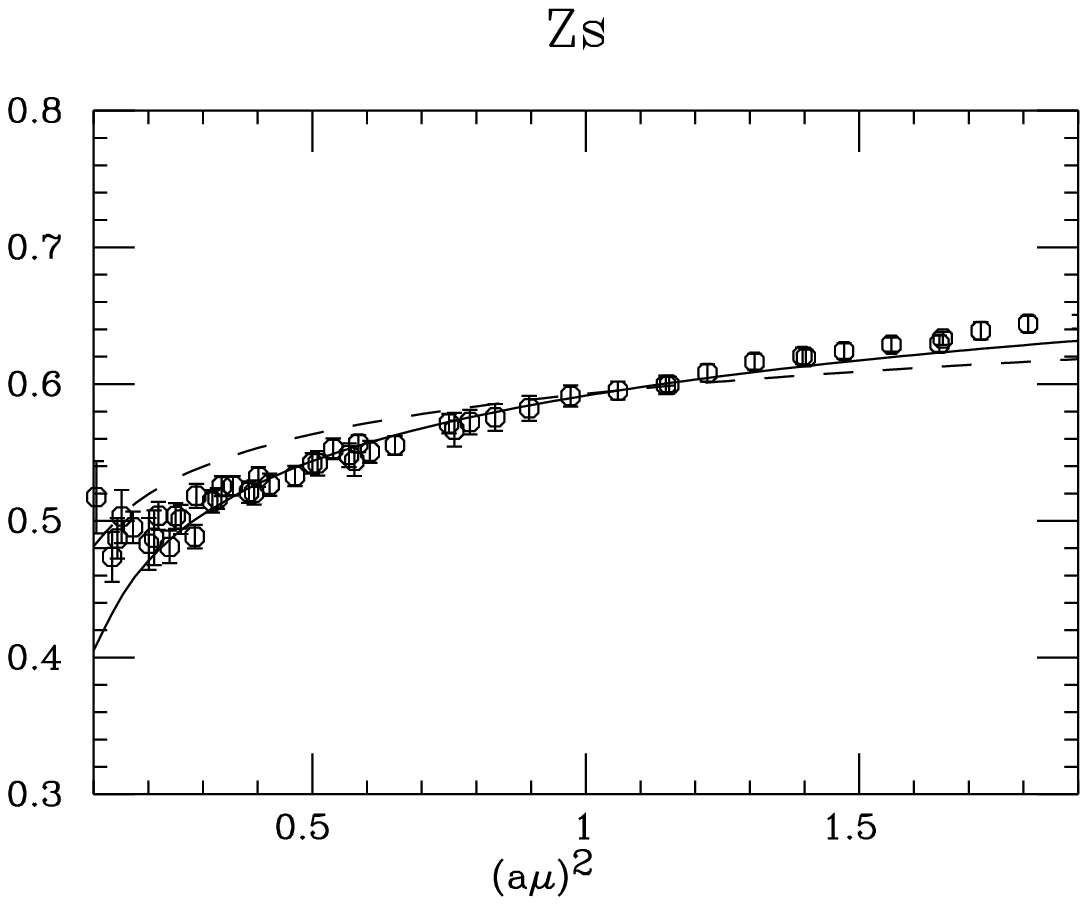}{100mm}
\vspace{3.0truecm}
\caption[]{\it $Z_S$ obtained by using the NPM as a function of the
scale $\mu^2 a^2$. The dashed and solid curves represent the solutions
of the renormalization 
group equations  at the leading  and next-to-leading  order, respectively.}
\protect\label{fig:zs}
\end{figure}

 
As a consistency check of our results, we have also studied the $\mu$ 
dependence
of $Z_S$. In fig.~\ref{fig:zs}
we show  the behaviour of $Z_S$ as a function of $\mu$
compared with the solution of the renormalization group equations
\beq Z_S(\mu) = Z_S(\mu_0) \left(\frac{\alpha_s(\mu)}{\alpha_s(\mu_0)}
\right)^{\gamma_0/2 \beta_0} \left( 1 +  \frac{\alpha_s(\mu)-
\alpha_s(\mu_0)}{4 \pi} J   \right) \ ,\label{eq:rgzs} \eeq
where, in Landau RI and with $n_f=0$~\cite{franco}, 
\bea \gamma_0 = - 8 \ , \quad \gamma_1 &=& -252 \ , \quad 
\beta_0 = 11  \ , \quad \beta_1 =  102 \nonumber \\
J&=& \frac{\gamma_1 \beta_0 - \gamma_0 \beta_1}{2 \beta^2_0} \ .
\label{eq:requa} \eea
%___________________________________________________________________________
%___________________________________________________________________________
\begin{table}
\begin{center}
\begin{tabular}{||c|c|c|c|c||} \hline\hline \multicolumn{5}{||c||}
{$\beta=6.2$} \\ \hline
 $\mu$ &  $Z_S$ (NPM)&  $Z_S$ (BPT)&  $Z_P$ (NPM)& $Z_P$ (BPT)\\ \hline 
$2.80$ GeV & 0.60(1)  & $0.66$ & $0.47(1)$ & $0.62$ \\
$2.00$ GeV & 0.55(1)  & $0.61$ & $0.43(1)$ & $0.57$ \\
 \hline \hline \multicolumn{5}{||c||}{$\beta=6.0$} \\\hline
 $\mu$ &  $Z_S$ (NPM)&  $Z_S$ (BPT)&  $Z_P$ (NPM)& $Z_P$ (BPT)\\ \hline 
$2.00$ GeV & 0.55(3)  & $0.63$ & $0.39(3)$ & $0.59$ \\
 \hline \hline\end{tabular}
\protect\label{tab:zs}
\caption{{\it $Z_S$ and $Z_P$ obtained by using  the NPM and boosted 
perturbation theory (BPT), at two typical reference scales, $\mu= 2.80$ GeV,
corresponding to $\mu^2  a^2 =1$ at $\beta=6.2$, and $\mu=2$ GeV which is 
the reference renormalization scale for the quark masses in lattice
calculations. We also used the values of $Z_A$ as obtained with the NPM, 
namely  $Z_A=0.793(5)$ and $Z_A=0.78(1)$ at $\beta=6.2$ and $6.0$ 
respectively.}}
\end{center}
\end{table}
By fitting the numerical results to eq.~(\ref{eq:rgzs}), with $\Lambda_{QCD}$
as free parameter, we get $\Lambda^{n_f=0}_{QCD} = 200 \pm 55$ MeV 
in good agreement with the determination of ref.~\cite{alpha}, $\Lambda ^{n_f=0}_{QCD} \sim 251\pm 21$ MeV. It is reassuring that eq.~(\ref{eq:rgzs})
fits rather well the behaviour of $Z_S(\mu)$ down to rather low values of
$\mu^2$. From the fit, and using $Z_P/Z_S=0.78(1)$, we find the results given
in table~1. We stress that the use of perturbation theory induces an error
of $10$--$30 \%$  in the  extraction of the values of the quark masses, 
depending whether one uses the vector or axial-vector Ward  identity (i.e. 
$Z_S$ or $Z_P$).
For completeness we also give the results at $\beta=6.0$.
More details on the analysis of the $Z$s obtained with the NPM 
can be found in ref.~\cite{mescia2}~\footnote{The amputated correlation 
function of the pseudoscalar density, between external off-shell quark states, 
is expected to receive in the chiral limit a non-perturbative contribution 
from the Goldstone pole~\cite{g3}. We have eliminated such a contribution 
and determined the renormalization constant $Z_P$ by evaluating the ratio 
$Z_P/Z_S$ at large values of $\mu^2$ and then calculating $Z_P$ as 
$(Z_P/Z_S) \cdot Z_S$.}. 
\par Eqs.~(\ref{eq:massav}) and (\ref{eq:massaa}) determine 
the bare masses $\tilde m$ and $ \tilde \rho$ as a function of the hopping
parameter $\kappa$. The physical values of $\kappa$, i.e. $\kappa_l$, 
$\kappa_s$ etc.,  
are fixed, together with the lattice spacing, using a certain number of 
physical conditions. 
We now explain the procedure followed in our analysis.
\par 
We fit  the mass of the vector meson to the expression
\beq a M_V = C + L (a M_{PS})^2 + Q (a M_{PS})^4 \ , \label{eq:fita}\eeq
where $M_{PS}$ is the pseudoscalar mass.  The cases with   $Q \neq 0$ or $Q=0$
are denoted as quadratic or linear fit respectively. By taking   $M_{PS}= 
m_\pi$ 
($M_{PS}= m_K$) 
and $M_V=m_\rho$ ($M_V=m_{K^*}$), where $m_\pi$ and $m_\rho$ are the
experimental numbers, we determine the value of the lattice spacing.
This determination, being  based on the physical spectrum  without
reference to any definition of the quark masses, is valid up to ${\cal O}
(a^2)$.
It has been called ``lattice-plane method" ($\rho$--$\pi$ or
$K^*$--$K$) in ref.~\cite{gimenez}.
To determine $\tilde m$ and $ \tilde \rho$, we must study another physical
quantity. The best (and most popular) choice is $M^2_{PS}$, since it
vanishes in the chiral limit  and for this reason is very sensitive to the
precise value of the quark masses. We fit $M^2_{PS}$ to
\bea a^2 M^2_{PS} &=& L_{PS} a (\tilde m_1 + \tilde m_2) +
Q_{PS} a^2 (\tilde m_1 + \tilde m_2)^2  \label{eq:mm}\ , \\
a^2 M^2_{PS} &=& L^\prime_{PS} a (\tilde \rho_1 + \tilde \rho_2) +
Q^\prime_{PS} a^2 (\tilde \rho_1 + \tilde \rho_2)^2 \label{eq:rr}\ , \eea
depending whether we use the vector or axial-vector WI.
In the above equations, with the values of $a$ fixed from
eq.~(\ref{eq:fita}) and by imposing
$M_{PS}=m_\pi$ or $m_K$,  we determine $\tilde m_i$  ($\tilde 
\rho_i$)~\footnote{ 
 Since we always work with degenerate quarks, we are unable to fit a term 
 $\propto 
(\tilde m_1 - \tilde m_2)^2$. In order to separate $\tilde m_l$
from $\tilde m_s$ (using $m_\pi$ and $m_K$),
we must,  however, distinguish the two quark masses.}.  For consistency, 
 quadratic or linear fits should be used both in eq.~(\ref{eq:fita})
and eq.~(\ref{eq:mm}) or (\ref{eq:rr}). We have  also determined the strange
quark mass from the vector meson mass, by fitting $M_V$ to
\bea a M_V &=& C_V +  L_V a (\tilde m_1 + \tilde m_2) +
Q_V a^2 (\tilde m_1 + \tilde m_2)^2 \ , \label{eq:dit1} \\
a M_V &=& C_V +  L^\prime_V a (\tilde \rho_1 + \tilde \rho_2) +
Q^\prime_V a^2 (\tilde \rho_1 + \tilde \rho_2)^2 \ , \label{eq:ditroppo} \eea
and using the experimental value of  $m_\phi$. 
\par 
>From the analysis of the results obtained
using quadratic and linear fits, we reached the following conclusions
\begin{itemize} 
\item As already observed in ref.~\cite{petronzio},
the coefficient $Q_{PS}$ is  large and positive, 
$Q_{PS}=1.54(14)$. It is very difficult to explain  the sign of $Q_{PS}$
as a consequence  of the use of  perturbation theory  for $b_m$: 
such an explanation would require the perturbative value of $b_m$ to be wrong 
both in size 
and in {\it sign}.  This is very
unlikely (and to our knowledge never happened to be the case).
Thus, we believe that the positive curvature $Q_{PS}$ is a physical effect 
(at least in
the quenched approximation).
\item In the range of masses used in our calculation, corresponding to the
values of $\kappa=0.1333$, $0.1344$, $0.1349$ and $0.1352$ (from heavier to 
lighter), 
the  quadratic fits on all 
the four quarks  give consistent results and statistical errors
for the quark masses. Moreover, the value of $\kappa_{critic}$ is
in agreement with that obtained form the axial-vector WI.
\item With our data, instead,  a linear fit of the pseudoscalar mass
on all the four values of $\kappa$ gives a value of $\kappa_{critic}$ 
incompatible with
that obtained from the axial-vector WI. This confirms 
that we need a quadratic fit for the pseudoscalar mass. 
\item Within the accuracy of our results, the coefficient 
$Q=-2.2 \pm 1.5$ ($Q_V=-2.1 \pm 2.5$)
is compatible with zero. Nevertheless, by using the linear or quadratic
fit to $M_V$, the central value  of the lattice spacing, and the 
value of the strange quark mass, vary by about $10 \%$.
\item If we assume that the quadratic correction to $M_V$ is negligible,
i.e. if we take $Q=0$ but  fit quadratically $M^2_{PS}$ (as done in 
ref.~\cite{petronzio}),
we obtain $a^{-1}=2.59(12)$ GeV (using $m_\rho$--$m_\pi$)
and $a \tilde m_s= 0.031(3)$ in excellent agreement with 
refs.~\cite{gockeler,petronzio}. 
\item With a quadratic fit for both $M_V$ and $M_{PS}^2$, however, 
we obtain a value of the inverse lattice spacing
which is sensibly higher than $2.59$ GeV, i.e. $a^{-1}=2.84(32)$. Although
the two results look compatible within the errors, we believe that the shift 
is a
systematic effect, since the two values of $a^{-1}$ are determined on the same
 set of
configurations.  Correspondingly, the values of the quark masses are also
modified. In the quadratic case, we get $a \tilde m_s=0.026(5)$. We stress 
again that
the shift of $a \tilde m_s$ toward smaller values, although compatible with 
$0.031(3)$,
is a systematic effect. 
\item  It is well known, see for example refs.~\cite{gupta,japs,gimenez},
that the  strange quark mass extracted using $m_K$ differs from that obtained 
from 
$m_\phi$ by about $15  \%$.  Moreover, the values of the inverse lattice 
spacing
determined from $m_\rho$--$m_\pi$ or  $m_K$--$m_{K^*}$
are slightly different (by about $100$ MeV). These effects were found  
irrespectively of the action  (Wilson, tree-level improved or
non-perturbatively improved) used in the numerical calculations and their 
origin 
can be traced from the fact
that the slope $L$ in eq.~(\ref{eq:fita}) is smaller than its experimental
value. With our data, we find the same effects if we use the  linear fit
for $M_V$ (with a difference of about $10$--$15 \%$ for $a \tilde m_s$, for 
example).
With  quadratic fits, we find  two nice features: on the one hand 
the difference in the value of $a^{-1}$ from $m_\rho$--$m_\pi$ and 
$m_{K^*}$--$m_K$
is strongly reduced; on the other, we get about the same strange quark mass
from $m_K$ and $m_\phi$. \par  Unfortunately, within  our
precision, we are unable to fix $Q$ well enough.  Nevertheless, at the price 
of increasing
the statistical error, we will take as best estimates
of the quark masses  those obtained by using quadratic fits for both
$M_V$ and $M^2_{PS}$.
Since this is at present one of the largest sources of uncertainty,  we 
believe that 
a precise measurement of the quadratic term 
in the dependence of $M_V$ on $M_{PS}$  is crucial  to achieve an accurate 
determination of  the mass of the strange quark. 
\end{itemize}
\par 
\begin{table}
\begin{center}
\begin{tabular}{||ccc|c|c|c|c||}
\hline
\hline \multicolumn{7}{||c||}{$\beta=6.2$} \\ \hline
Method & $a^{-1}$ & Input & $\tilde m_l a $ & $m^{\mbox{\scriptsize{RI}}}_l$ & 
  $\tilde \rho_l a $ & $m^{\mbox{\scriptsize{RI}}}_l$ \\
 & (GeV) & & & (MeV) & & (MeV) \\ \hline
$\rho$--$\pi$ L4Q & $2.59(12)$& $m_\pi$ & 0.00130(11)& 6.1(3) &0.00127(11) &
  6.0(3) \\ \hline
$K^*$--$K$    L4Q & $2.69(12)$& $m_\pi$ & 0.00120(10)& 5.9(3) &0.00117(10) &
  5.8(3) \\ \hline
$\rho$--$\pi$ QQ  & $2.84(32)$& $m_\pi$ & 0.00108(23)& 5.6(6) &0.00105(24) &
  5.5(6) \\ \hline
$K^*$--$K$    QQ  & $2.83(25)$& $m_\pi$ & 0.00109(18)& 5.6(5) &0.00106(18) &
  5.5(5) \\ 
 \hline \hline
Method & $a^{-1}$  &Input  & $\tilde m_s a $ & $m^{\mbox{\scriptsize{RI}}}_s$ &
 $\tilde \rho_s a $ & $m^{\mbox{\scriptsize{RI}}}_s$ \\
& (GeV) & & & (MeV) & & (MeV) \\ \hline
$\rho$--$\pi$ L4Q & $2.59(12)$& 
        $m_K$ & 0.0312(26) & 147(6) & 0.0313(28) & 149(7) \\ 
 & & $m_\phi$ & 0.0363(35) & 171(9) & 0.0379(40) & 181(12) \\ \hline
$K^*$--$K$    L4Q & $2.69(12)$& 
        $m_K$ & 0.0290(23) & 142(6) & 0.0291(24) & 144(6) \\ 
 & & $m_\phi$ & 0.0319(24) & 156(5) & 0.0331(27) & 164(7) \\ \hline
$\rho$--$\pi$  QQ & $2.84(32)$& 
        $m_K$ & 0.0261(55) & 135(14) & 0.0261(58) & 137(16) \\
& &$m_{\phi}$ & 0.0260(84) & 134(29) & 0.0265(94) & 139(35) \\ \hline
$K^*$--$K$ QQ & $2.83(25)$  &   
        $m_K$ & 0.0263(42) & 135(11) & 0.0263(45) & 137(12) \\
& &$m_{\phi}$ & 0.0264(58) & 136(19) & 0.0270(66) & 141(23) \\
\hline\hline
\end{tabular}
\protect\label{tab:qmasses}
\caption{{\it Improved bare masses of the light ($\tilde m_l a=
(\tilde m_u a + \tilde
m_d a)/2$ and $\tilde \rho_l a=(\tilde \rho_u a + \tilde
\rho_d a)/2$) and strange ($\tilde m_s a$ and $\tilde \rho_s a$) 
quarks in lattice units. The improvement coefficients $b_m$, $b_A$ and
$b_P$ necessary to obtain $\tilde m_l$, $\tilde \rho_l$,  etc.,
have been taken from boosted perturbation theory. 
The renormalized quark masses in the RI scheme
at a renormalization scale $\mu=2$ GeV, 
obtained using the calibration of the lattice spacing given in this table, 
and the  non-perturbative values of the $Z$s from tab.~1, 
are also given.
We use the values of the inverse lattice spacing
obtained from $m_\rho$-$m_\pi$ (denoted as $\rho$--$\pi$)   and from 
$m_{K^*}$-$m_K$ (denoted as $K^*$--$K$). L4Q denotes the same fitting procedure
as in ref.~\protect\cite{petronzio}: a linear fit for $M_V$, over the
four quark masses, combined with a quadratic fit for $M^2_{PS}$. These results
can be directly compared to those of refs.~\protect\cite{petronzio}.
QQ indicates that quadratic fits were used  for both $M_V$ and
$M^2_{PS}$.}}
\end{center}
\end{table}
In table~2, we present a rather extended set of results for the
  bare and renormalized quark masses, obtained from the vector and axial-vector
Ward identities, with the  $b$s computed in boosted perturbation theory. 
The renormalized masses, in the RI scheme  at $\mu = 2$ GeV, have been obtained
from the bare ones using the renormalization constants of tab.~1. 
We see from this table that, by using quadratic fits, denoted as QQ in the 
table,
and non-perturbative $Z$s,  we obtain  very  consistent results for the
two calibrations of the lattice spacing, the masses from the
vector or axial-vector WIs and the quark mass extracted from $m_K$ or $m_\phi$.
In particular, we  stress  the excellent agreement between the values
of the quark masses obtained from the vector and axial-vector WI.
This agreement is only possible  if one uses $Z_S$ and $Z_P$ computed
non-perturbatively. This was first noticed  in ref.~\cite{gimenez}. 
A sizeable difference remains,  instead,  if one uses boosted 
perturbation theory, see also \cite{gockeler}. For example, 
with the perturbative values of $Z_S$ and $Z_P$, we obtain,
at $\mu=2$ GeV,
$m^{\mbox{\scriptsize{RI}}}_s=118(12)$ MeV (from $\tilde m_s$) and
$m^{\mbox{\scriptsize{RI}}}_s=101(11)$ MeV (from $\tilde \rho_s$)
instead of $135(14)$ MeV and $138(16)$ MeV, respectively. 
\par From the results of the table, we extract our best estimate of the light
quark masses in the RI scheme and  at $\mu=2$ GeV  
(in the following all the results for the
running masses refer to $\mu=2$ GeV)
\beq m^{RI}_l = 5.6(5) \, \mbox{MeV} \quad 
     m^{RI}_s = 138(15) \, \mbox{MeV} \label{eq:mris}\ . \eeq
Note that, to obtain the results in eq.~(\ref{eq:mris}), 
we never used perturbation theory but for the values of $b_m$, $b_A$ and 
$b_P$. However, as discussed above, this is expected to be a source of 
negligible uncertainty in the determination of light quark masses.
\section{$\overline{MS}$ and RGI masses} 
Perturbation theory only enters  if we want to convert the results of  the RI
scheme into the $\overline{MS}$ scheme, which has been adopted as the standard
one in the literature (both in lattice and QCD sum rule calculations). 
Although this is not necessary (and we believe that $m^{\scriptsize{RGI}}$  
is a more convenient definition, see below), for comparison 
with other determinations, we also give the quark masses in  $\overline{MS}$.
These  are found using the relation
\beq m^{\scriptsize{\overline{MS}}}(\mu)= R_m \,
m^{\scriptsize{RI}}(\mu) \label{eq:conversion} \ , \eeq 
where, following the notation of ref.~\cite{franco} 
\begin{equation}
\label{eq:rm}
R_m = 
1+ \frac{\alpha_s(\mu)}{\left( 4\pi \right) }\left( Z_m^{\scriptsize{RI}}
\right)_0^{(1)}
+\frac{\alpha_s^2(\mu)}{\left( 4\pi \right) ^2}\left( Z_m^{\scriptsize{RI}}
\right) _0^{(2)} 
+ \ldots \,
\end{equation}
with 
\begin{eqnarray}
\label{eq:zmri}
\left( Z^{\scriptsize{RI}}_m \right)^{(1)}_{0} & = & 
 -8  {{\left(N_c^2 - 1 \right) }\over {4\,N_c}} \ , 
 \nonumber\\
\left( Z^{\scriptsize{RI}}_m \right)^{(2)}_{0} & = & 
  {{\left(N_c^2 -1 \right) }\over {96\,N_c^2}} \,
  \left( -75  - 2645\,N_c^2
 + 288\,\zeta_3 + 576\,N_c^2\,\zeta_3 + 356\,N_c\,n_f  \right) 
\end{eqnarray}
In the above equation, $\zeta _3=1.20206\ldots$ is
the Riemann zeta function. 
Since all previous results have been obtained at the NLO, i.e. by ignoring the
corrections due to $\left( Z^{\scriptsize{RI}}_m \right)^{(2)}_{0}$,
we give the results both at the NLO and at the N$^2$LO  order.
In the numerical evaluation of $R_m$ we have used $n_f=4$. The reason for this
choice, in spite of the quenched approximation  adopted in our simulation,
is the following. The mass $m^{\scriptsize{RI}}(\mu)$ is to be interpreted 
as the mass in the continuum which includes a systematic (unknown) error 
coming from the quenched approximation. Thus, $m^{\scriptsize{RI}}(\mu)$  
is the estimate of the physical
value of the quark mass at the scale $\mu=2$ GeV, at which $n_f=4$.
By fixing in both cases $\alpha_s(M_Z)=0.118$, which corresponds
to $\alpha^{\scriptsize{NLO}}_s(\mu=2$ GeV$)=0.296$ and
$\alpha^{\scriptsize{N}^2\scriptsize{LO}}_s(\mu=2$ GeV$)=0.300$, we obtain
$R_m^{\scriptsize{NLO}}=0.874$ and $R_m^{\scriptsize{N}^2\scriptsize{LO}}=
0.804$
which give
\bea 
m^{\scriptsize{NLO \ \overline{MS}}}_l&=& 4.9(4) \, \mbox{MeV} \ , \quad 
m^{\scriptsize{NLO \ \overline{MS}}}_s = 121(13) \, \mbox{MeV}\nonumber \\ 
m^{\scriptsize{N}^2\scriptsize{LO \ \overline{MS}}}_l&=& 
4.5(4) \, \mbox{MeV} 
\ , \quad m^{\scriptsize{N}^2\scriptsize{LO \ \overline{MS}}}_s= 
111(12) \, \mbox{MeV}\label{eq:mms}  \ . \eea 
Note that {\it by using BPT, we would have been obtained},
for example, $m^{\scriptsize{NLO \ \overline{MS}}}_l= 
4.3$ MeV {\it and} $m^{\scriptsize{NLO \ \overline{MS}}}_s=98 (11)$ MeV
(from the vector WI, which is the most favorable case for BPT).
\par We now compare our results to those of refs.~\cite{gockeler} and
\cite{petronzio}.  For  $\mu=2$ GeV, we
rescale their results with the ratios of the
perturbative $Z$s to the non-perturbative ones, at the
corresponding values of $\beta$.  In this way, using the results
of ref.~\cite{gockeler}, we get 
$m^{\scriptsize{NLO \ \overline{MS}}}_s=126 (5)$ MeV
from the vector WI and 
$m^{\scriptsize{NLO \ \overline{MS}}}_s=138 (3)$ MeV
from the axial-vector WI and, from ref.~\cite{petronzio},
$m^{\scriptsize{NLO \ \overline{MS}}}_s=127
(15)$ MeV from the vector WI. These numbers are in good agreement with 
the NLO result given in  eq.~(\ref{eq:mms}). 
\par We also analyzed the  data of our test run at $\beta=6.0$. In this
case we obtain  rather larger values:
$m^{\scriptsize{NLO \ \overline{MS}}}_s=145(7)$ MeV
from the vector WI, using quadratic fits 
($m^{\scriptsize{NLO \ \overline{MS}}}_s=145(14)$ MeV
from the axial WI). 
With linear fits on the lightest
quarks, corresponding to $\kappa=0.1324$, $0.1333$ and $0.1342$, we obtain
a result much closer to that obtained at $\beta=6.2$, i.e. 
$m^{\scriptsize{NLO \ \overline{MS}}}_s=126 (6)$ MeV.
As a further check, we have fitted the results of ref.~\cite{gockeler}
with our programs (with quadratic fits), obtaining
$m^{\scriptsize{NLO \ \overline{MS}}}_s=131$ MeV~\footnote{ 
It is not possible to extract an error in this case.}.
The problem is  that our accuracy at $\beta=6.0$ is rather poor, 
and we attribute
the discrepancy between the results with quadratic fits and all the others
to the fact that we do not control, in this case, the quadratic corrections.
These are affected by   some differences
with the more precise results of ref~\cite{gockeler}, 
which  we have found for the values of the meson masses for the lightest quarks.
\par A originally  discussed in ref.~\cite{paco} for lattice calculations, 
and more recently stressed by the Alpha collaboration, a more convenient 
definition of the quark mass  is  $m^{\scriptsize{RGI}}$  which, analogously 
to the RGI  $B$-parameter,  $\hat B_K$, is  also renormalization scheme
and scale invariant. Moreover, $m^{\scriptsize{RGI}}$ can 
be directly related to the mass parameters  appearing in the
fundamental Lagrangian  of grand unified theories, at the unification scale.
The relation between $m^{\scriptsize{RGI}}$ and $m^{\scriptsize{RI}}(\mu)$  is
given by~\cite{franco}
\bea
m^{\scriptsize{RGI}}_{n_f=4} &\equiv& R_{RGI} \, 
m^{\scriptsize{RI}}\left(\mu\right)
 = \alpha_s\left(\mu\right)^{-12/25}
\left[ 1 - \frac{\alpha_s\left(\mu\right)}{4\pi}\left( \frac{17606}{1875} 
\right) \right. \\
&\phantom{x}& \left. - \frac{\alpha_s^2 \left(\mu\right)}{\left( 4\pi \right)
^2 }
\left( \frac{3819632767}{21093750}-\frac{952}{15}\zeta _3 
 \right) \right] 
m^{\scriptsize{RI}}\left(\mu\right)  \  . \label{eq:rgim} \eea
With the same values of $\alpha_s$ given above, we find 
$R_{RGI}^{\scriptsize{NLO}}=1.40$ and  
$R_{RGI}^{\scriptsize{N}^2\scriptsize{LO}}=1.28$, corresponding to
\bea \hat m^{\scriptsize{NLO}}_l&=& 
7.8(7) \, \mbox{MeV}
\ , \quad \hat m^{\scriptsize{NLO}}_s= 
193 (21) \, \mbox{MeV}\nonumber \\ 
\hat m^{\scriptsize{N}^2\scriptsize{LO}}_l&=& 
7.2(6) \, \mbox{MeV} 
\ , \quad \hat m^{\scriptsize{N}^2\scriptsize{LO}}_s= 
177(19) \, \mbox{MeV}\label{eq:mmms}  \ . \eea 
\section{Conclusions}
Using the ${\cal O}(a^2)$  improved Alpha action,
we have computed the light-quark masses renormalized   non-perturbatively. 
The NPM used in this study removes the discrepancies, 
observed in previous calculations, between the results obtained using the 
vector  and the 
axial-vector Ward identities. It also gives values of the quark masses larger
than those obtained by computing the renormalization constants with (boosted)
perturbation theory.   We  use perturbation theory only to translate our
results either to the  $\overline{MS}$ scheme, which has been adopted in most 
of the determinations of the quark masses, or to our preferred definition
which is the RGI mass. Using the results of ref.~\cite{franco} this
can be done at $N^2LO$ accuracy.
\par A further reduction of the systematic errors would require the 
non-perturbative determination of the $b$s (although we believe that this 
uncertainty
is,  for light quarks, by far smaller than the others); the evaluation of 
$Z_S(\mu)$ and $Z_P(\mu)$ at larger physical scales, $\mu$, and smaller values
of $\mu a$, a  fit of the renormalized  masses in $a^2$ (in order to remove
the residual  discretization errors) and, of course, precise unquenched 
calculations.
We also found that a precise determination of the quadratic dependence of the 
vector meson mass on the quark mass is important to understand the 
differences in  $\tilde m_s$  determined  using different  mesons ($\sim 20$ 
MeV).

\section*{Acknowledgements}
V.L. and G.M. acknowledge the M.U.R.S.T. and the INFN for partial support. 
We are grateful to A. Le Yaouanc for useful and interesting discussions.

\end{document}